# Beyond Vaccination Rates: A Synthetic Random Proxy Metric of Total SARS-CoV-2 Immunity Seroprevalence in the Community

Yajuan Si[a], Leonard Covello[b], Siquan Wang[c], Theodore Covello[d], and Andrew Gelman[e]

[a] Institute for Social Research, University of Michigan, Ann Arbor, MI

[b] Community Hospital, Munster, Indiana

[c] Department of Biostatistics, Columbia University, New York, NY

[d] Department of Biology, University of Chicago, Chicago, IL

[e] Departments of Statistics and Political Science, Columbia University, New York, NY

Corresponding author: Yajuan Si, Email: yajuan@umich.edu, Telephone: 734-764-6935,

Survey Research Center, Institute for Social Research, University of Michigan,

ISR 4014, 426 Thompson St, Ann Arbor, MI 40104

Running head: A Proxy Metric for Total SARS-CoV-2 Immunity Seroprevalence

Conflict of interest: none declared.

Funding support: This study was supported by the National Science Foundation and National Institutes of Health.

**Reproducibility:** Codes are publicly available at https://github.com/yajuansi-sophie/covid19-mrp. The data are confidential and cannot be released to the public.

Human subjects research: The study of secondary data analysis does not involve human subjects research and is exempted from the independent review board.

# Beyond Vaccination Rates: A Synthetic Random Proxy Metric of Total SARS-CoV-2 Immunity Seroprevalence in the Community

ABSTRACT

Explicit knowledge of total community-level immune seroprevalence is critical to developing policies to mitigate the social and clinical impact of SARS-CoV-2. Publicly available vaccination data are frequently cited as a proxy for population immunity, but this metric ignores the effects of naturally-acquired immunity, which varies broadly throughout the country and world. Without broad or random sampling of the population, accurate measurement of persistent immunity post natural infection is generally unavailable. To enable tracking of both naturally-acquired and vaccine-induced immunity, we set up a synthetic random proxy based on routine hospital testing for estimating total Immunoglobulin G (IgG) prevalence in the sampled community. Our approach analyzes viral IgG testing data of asymptomatic patients who present for elective procedures within a hospital system. We apply multilevel regression and poststratification to adjust for demographic and geographic discrepancies between the sample and the community population. We then apply state-based vaccination data to categorize immune status as driven by natural infection or by vaccine. We have validated the model using verified clinical metrics of viral and symptomatic disease incidence to show the expected biological correlation of these entities with the timing, rate, and magnitude of seroprevalence. In mid-July 2021, the estimated immunity level was 74% with the administered vaccination rate of 45% in the two counties. The metric improves real-time understanding of immunity to COVID-19 as it evolves and the coordination of policy responses to the disease, toward an inexpensive and easily operational surveillance system that transcends the limits of vaccination datasets alone.

INTRODUCTION

The COVID-19 pandemic has drastically affected everyday life for nearly two years, claiming millions of lives and shuttering businesses worldwide. Even as we have made progress with the ongoing implementation of vaccines, new challenges continue to arise. Vaccination numbers remain stubbornly limited and among some demographics, there is a hardened resistance to participate. Furthermore, reinfection with SARS-CoV-2 and vaccine breakthrough have both been observed, and new potentially antibody-resistant virus variants are likely to continue to evolve.

To effectively develop policies to address an epidemic such as COVID-19, we require explicit knowledge of the level of immunity in any given community, state, or across the nation. Understanding how many people have functional immunity, whether by vaccine or through natural acquisition, allows for a better prediction of how resistant a community may be to severe infection. This information would inform mitigation strategies and help to manage social and business interactions. State vaccination records are followed intensely in the media and scientific community (1), but only tell part of the story. There is an explicit need to comprehend the place of naturally acquired immunity as well.

Though incompletely informative, the presence of antibody Immunoglobulin G (IgG) to COVID-19 is an important indicator of immunity within a given population. Assays for IgG both to the spike protein (IgG S) and to the nucleocapsid protein (IgG N) are broadly available commercially.

Both antibodies will be present in individuals who have acquired natural immunity, while vaccinated individuals will only have the IgG S antibody, since the genetic material encoding the S protein, and not the virus itself, is what is administered through the various vaccines. As a result, not only can such antibody assays detect immunity, but they can also potentially distinguish between these two forms of immunity. Systematic trending of both forms of seropositivity provides multiple benefits for understanding the presence and evolution of immune status within the community, both naturally and vaccine-acquired.

Ideally, any given population could be randomly sampled serially, and the shifting antibody levels trended. Although such sampling has been done to assess IgG prevalence nationwide at a given moment, this method is cumbersome, expensive, and impractical for tracking levels over time and in various jurisdictions. In our hospital system, we have been tracking the level of IgG N in our community since May 2020 and both IgG N and IgG S since February 2021 using a method that is created to function as a proxy for synthetic random sampling. By serially assaying patients for IgG N as part of their routine presurgical workup, we generate a demographically stable serial sample in high numbers, at low cost, and repeatable over time, and then normalize the demographics of the sample to those of the community at large using multilevel regression and poststratification (MRP) (2). In doing so, we generate a satisfactory approximation for random sampling, as we are able to validate in the analysis that follows. Our method can be easily adapted to incorporate data from multiple hospital systems to provide robust population inferences. We advocate for the broad adoption of this metric to inform policymakers about immunity levels in their respective jurisdictions and are engaged in building a server-based utility to facilitate the data and result management.

DATA

All preoperative patients in our three-hospital system (Community Hospital, St. Catherine Hospital, and St. Mary Medical Center) were subjected to surgical risk evaluation according to accepted American Society of Anesthesiology standards. A subset of these patients was deemed to benefit from preoperative blood tests of whatever variety, as indicated by their age or health characteristics or by the intensity of the planned surgery. Since May 1, 2020, all such patients already needing preoperative blood testing were also tested for the presence of IgG N using a commercially available qualitative test (Roche). The actual threshold titer for this assay was chosen relatively high in order to improve specificity; SARS-CoV-2 IgG assays in general suffered from cross-reactivity to the human endemic coronaviruses, whose incidence was quite high throughout the U.S. population (3). At the chosen threshold, such cross-reactivity was virtually eliminated. Therefore, the results were expected to reflect extremely high specificity of identifying actual humoral immunity to COVID-19 (at or near 100% per vendor internal analysis), while sacrificing some sensitivity to lower titers of antibody in any given individual. The data were thus best viewed as a lower bound of IgG N positivity in the sample population.

With the advent of SARS-CoV-2 spike protein (S) mRNA-based vaccines in our community, we recognized the apparent importance of accounting for IgG S as well as it evolved in our vaccinated population. To that end, as of February 16, 2021, our protocol was modified to obtain both IgG N and IgG S results for the target population. The IgG S test was a semiquantitative ECLIA based assay that quantifies IgG S titers over a 300-fold range above a threshold minimum (Elecsys Anti-SARS-CoV-2 S). Again, the analytical specificities were at or near 100% for this assay.

METHODS

As in our previous study of SARS-CoV-2 RNA acute incidence (4), we recognize that the sample population differs from the community demographically — it is older, whiter, and sicker — and, therefore, any inference about the community incidence of SARS-CoV-2 IgG will require demographic normalization of the outputs. MRP has an established history of accounting for such demographic decomposition in the social science literature (5-7). With this goal in mind, our data were subjected to MRP according to our previously described statistical construct (4). That modification was rigorously validated statistically in our earlier study, and the output was deemed a quality proxy for true random sampling of the incidence of IgG to SARS-CoV-2.

In order to test the validity of our hypothesis, we need to first understand the clinical behavior of acute SARS-CoV-2 RNA incidence with respect to the timing of asymptomatic/symptomatic COVID-19 and, further, with respect to the timing and magnitude of the rise, plateau, and decay of SARS-CoV-2 IgG prevalence. Fortunately, these characteristics are already well-described in the literature (e.g., 8). Peak positivity in SARS-CoV-2 from pharyngeal samples was estimated to occur at a mean of about five days post viral transmission from infected host to new recipient for the variants extant at the time of our study. Moderate-to-severe symptoms, should they occur, were then expected to arise approximately a week later. An infected patient is then expected to develop IgG to the virus about a week after the onset of symptoms, reaching peak levels a few weeks later. Given these established observations, we employ valid clinical metrics for both positive RNA status and symptomatic disease incidence measured by emergency department (ED) visits and compare these proxies to our MRP modified IgG prevalence data to see if the above biological behavior is respected.

To that end, we first employed the same model for acute viral RNA incidence from pharyngeal samples and estimated the polymerase chain reaction (PCR) prevalence as that in Covello et al. (2021). The earlier study found that state-based positive case numbers and positivity rates substantially lagged any reasonable estimate of viral incidence (4). Instead, we demonstrated that a synthetic random proxy sampling of asymptomatic individuals gave the earliest and most reliable evidence of RNA positivity.

IgG N (and later, both N and S) positivity data were analyzed statistically and graphically. IgG N positivity in particular—as a marker of naturally acquired immunity—was compared with trends in ED visits in our community (Lake and Porter Counties, Indiana) and in asymptomatic MRP normalized RNA positivity. These latter metrics were deemed reasonable proxy measures for moderate-to-severe symptomatic COVID-19 and acute viral transmission, respectively, based on the known biology of the virus and our earlier analysis (4, 8). State based positivity rates, positive case numbers, and hospitalizations were rejected as proxies: the former due to its demonstrated poor reliability in our previous study and elsewhere (9), the latter two due to their demonstrated delay relative to the ED data in our study and on our state website (10).

We used a Bayesian approach accounting for unknown sensitivity and specificity (11) and applied MRP to testing records for population representation, here using the following adjustment variables: biological sex, age (0-17, 18-34, 35-64, 65-74, and 75+), race (Black, white, and other), and county (Lake and Porter). MRP has two key steps: first, fit a multilevel model for the prevalence with the adjustment variables based on the testing data; next, poststratify using the population distribution of the adjustment variables, yielding prevalence estimates in the target population. The statistical details are provided in the Appendix.

|  | Asymptomatic | | | Hospital | Community |
| --- | --- | --- | --- | --- | --- |
|  | PCR | IgG N | IgG S/N |  |  |
| Size | 39952 | 5065 | 2621 | 35838 | 654890 |
| Female, % | 59 | 60 | 61 | 57 | 51 |
| Male, % | 41 | 40 | 39 | 43 | 49 |
| Age0-17, % | 3 | 0.5 | 0.3 | 8.7 | 24 |
| Age18-34, % | 10 | 8 | 6.3 | 12 | 21 |
| Age35-64, % | 46 | 49 | 46 | 30 | 40 |
| Age65-74, % | 24 | 25 | 26 | 20 | 9 |
| Age75+, % | 17 | 17 | 21 | 29 | 6.6 |
| White, % | 72 | 73 | 73 | 65 | 69 |
| Black, % | 14 | 13 | 13 | 19 | 19 |
| Other, % | 14 | 14 | 14 | 16 | 12 |
| Lake, % | 84 | 91 | 92 | 88 | 74 |
| Porter, % | 16 | 9 | 7.7 | 12 | 26 |

Table 1. *Summary of Polymerase Chain Reaction (PCR), Immunoglobulin G (IgG) to the nucleocapsid protein (IgG N), and IgG to the spike protein (IgG S) test results and sociodemographics.*

RESULTS

From May 1, 2020 through July 12, 2021, our hospital system enrolled 39,952 asymptomatic patients who presented for elective procedures in this study. From May 1, 2020 through July 12, 2021, a subset of 5,065 qualitative IgG N assays (Roche Elecsys Cobas) were performed, selection based merely on the need for preprocedural blood testing. From February 16, 2021 to date, both the above IgG N assay and the semiquantitative IgG S assay (Roche Elecsys Cobas) were performed, comprising a further 2,621 patients.

Table 1 presents the descriptive summary sociodemographic distributions for asymptomatic patients and the population in the hospital system and the community. The overall prevalence is 0.7% for all PCR tests, 14% for IgG N only test from May 1, 2020 through July 12, 2021 and 64% for either IgG N or IgG S seropositivity from February 16, 2021 to date. The asymptomatic patients with IgG tests have similar sex and race decompositions to those with PCR tests but have a lower coverage of young or early adult patients (younger than 35 years old), the proportions of which are 6.6% for IgG S/N tests, 8.5% for IgG N tests, and 13% for PCR tests. The group with IgG S/N tests has more older adults (75 years and above, 21%), the vaccination of whom has initiated relatively early. As compared to the hospital system patients, asymptomatic patients with PCR/IgG tests tend to be female, middle-aged (35-64) or old (65-74), and white.

Given that IgG testing was only performed on those patients who would benefit from preoperative blood tests, the above skews are not surprising. For this reason, neither the hospital patients nor the asymptomatic patients serve as a precise representation of the community population, in particular with an under-coverage for young, male, and non-white residents. As it happens, these differences are not large; nonetheless, they are potential sources of error if not accounted for in our statistical model, and can also interfere with estimates of trends if the demographic breakdown of hospital patients varies over time. Furthermore, the county representation is unbalanced. Some patients are from south Cook County, Illinois, and are grouped into the Lake County as a proxy. Fortunately for our analysis, these contiguous communities have similar socioeconomic and ethnic demographics.

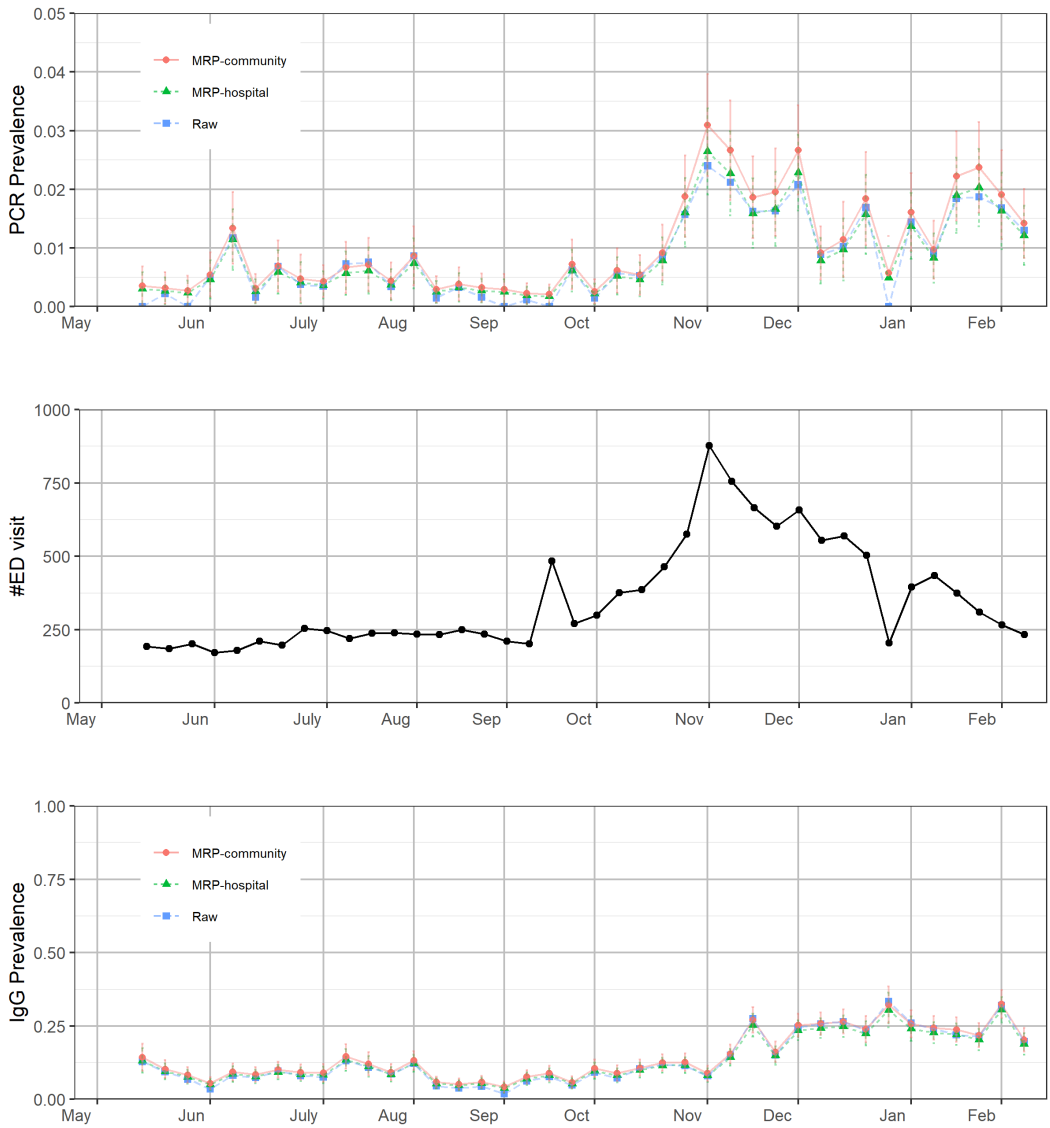

Figure 1. *Combined estimated Polymerase Chain Reaction (PCR) prevalence in the hospital system and community based on asymptomatic patients, COVID-related emergency department (ED) visits in Lake and Porter counties and estimated Immunoglobulin G (IgG) to the nucleocapsid protein (IgG N) prevalence till 02/15/2021. The error bars represent one standard deviation of uncertainty. The positions of the months on the x-axis correspond to the week of data containing the first of that month.*

The resulting outputs were then modified according to the previously-described MRP based demographic normalization. These results were subjected to the chosen clinical validation metrics. In Figure 1, we present three separate measures of viral behavior in our community. Figure 1A shows the MRP modified asymptomatic rates of viral RNA incidence in our sample. As a measure of clinical burden, we chose to follow trends in SARS-CoV-2 related ED visits for our catchment area, Lake and Porter Counties, Indiana. These data are presented graphically in Figure 1B. We then applied these established clinical metrics to assess the validity of our MRP modified IgG prevalence data, presented in parallel below (Figure 1C).

In northwest Indiana, there was a substantial outbreak of COVID-19 in March and April 2020, filling hospital intensive care unit beds and taxing the ventilator supply. As this study was initiated upon reopening of elective surgical services, May 1, 2020, the data presented here necessarily missed this trend clinically, but should be reflected in the development of a nonzero SARS-CoV-2 IgG prevalence. Indeed, Figure 1C shows that our MRP modified IgG prevalence hovered above 10% in May. We did not have an effective way to test the validity of this number, given the unreliable RNA testing availability at that time, but the result did not seem prima facie unreasonable to our group.

Following the progress of COVID 19 through the summer, it is evident from COVID-19 specific ED visits and MRP modified asymptomatic RNA incidence data (Figures 1A and 1B) that there was a very low prevalence of SARS-CoV-2 infection and therefore, very little anticipated accretion of IgG positivity during that interval. In fact, one observed an exceptionally slow decrease in IgG prevalence from May to August, eventually finding its nadir below 10% at the end of that period (Figure 1C). We believe that the very slow fade of IgG positivity during that interval reflects a slow decay of seropositivity beyond the minimal accretion of newly seropositive patients. As the acute infection data shows that new incidence is quite low, the decay in seropositivity is also quite slow, albeit modestly quicker than the accretion during this interval. The findings with those cohort studies served as a significant reassurance of its validity. We find more supporting evidence for validity during the fall surge. In the study area, both MRP modified asymptomatic RNA positivity and COVID-19 specific ED visits showed some evidence of an increase of trough numbers in mid-September. By early October, it had become apparent that a surge was evolving. A comparison of the peak in the RNA positivity or the number of ED visits to the slope and peak of the MRP IgG prevalence is revealing. We note a peak in both ED visits and RNA positivity in early November (Figure 1A and 1B). Peak incidence of acute infection should correlate with peak accretion of new IgG seroconversions; that is, the maximum *incidence* of viral infection seen in Figures 1A and 1B should correlate with the maximum *rate of increase* of IgG seropositivity in Figure 1C. From the figures, we find that the maximum incidence of the virus occurred in early November and the maximum slope of the IgG curve occurred in mid-November. This finding is consistent with the biological prediction that this largest cohort of viral positive patients will show maximum seroconversion a few weeks later (12), adding themselves at the highest rate to the IgG prevalence predicted in Figure 1C.

From early February to the conclusion of the study interval, ED visits remained near their summer 2020 lows. These findings are consistent with a host of studies that identify six months or longer persistence of IgG positivity in individuals, albeit at lower titers (13, 14). In the current population-based study, using a qualitative assay with a lower bound threshold, one would expect a prolonged loss of positivity in the assay as patients whose titers were at or near threshold fell into the negative range.

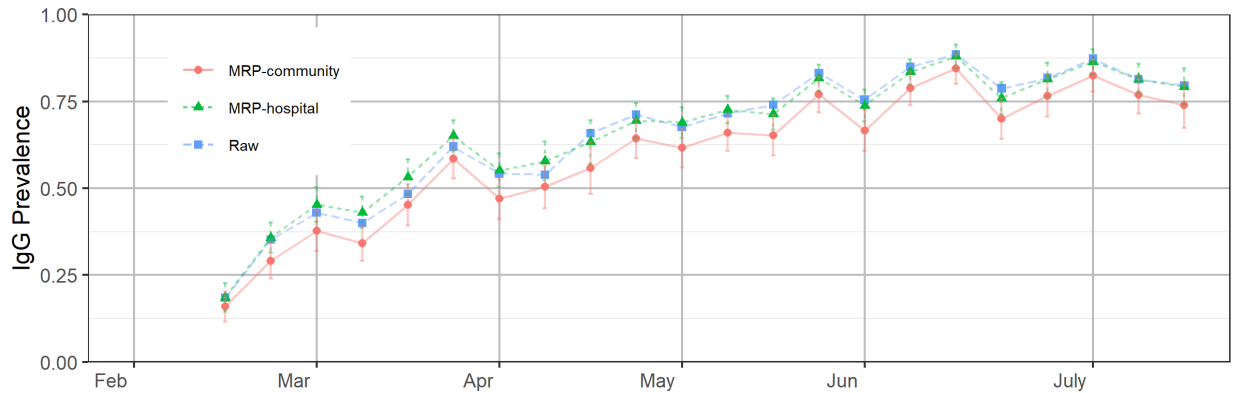

Figure 2. *Estimated combined Immunoglobulin G (IgG) to the nucleocapsid protein (IgG N) and the spike protein (IgG S) prevalence in the hospital system and the community based on asymptomatic patients after 02/15/2021. The error bars represent ±1 standard deviation of uncertainty. The positions of the months on the x-axis correspond to the week of data containing the first of that month.*

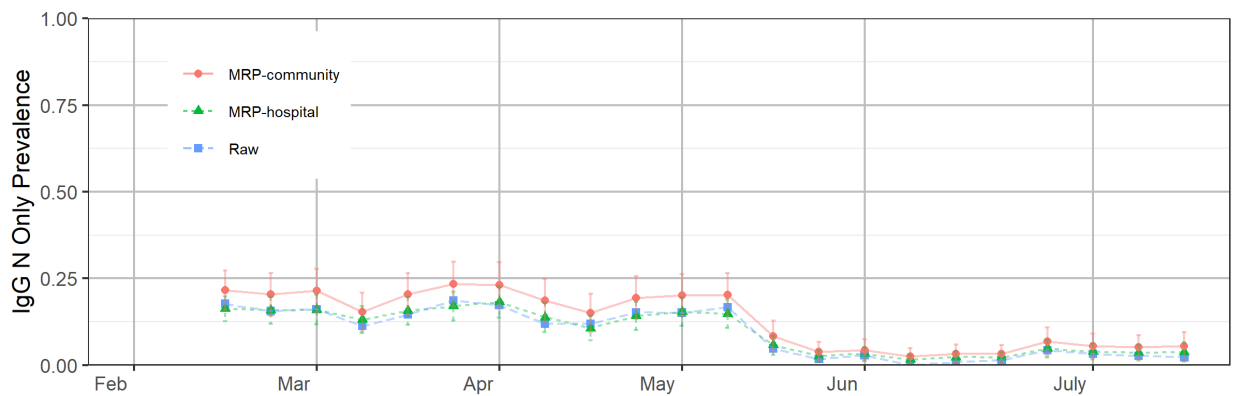

Figure 3. *Estimated combined Immunoglobulin N (IgG N) seropositive and Immunoglobulin S (IgG S) seronegative phenotype prevalence in the hospital system and the community based on asymptomatic patients after 02/15/2021. The error bars represent ±1 standard deviation of uncertainty. The positions of the months on the x-axis correspond to the week of data containing the first of that month.*

|  | IgG N/S positivity rate (standard error) | Vaccination rates (first/single dose administered) |
| --- | --- | --- |
| Overall | 0.74 (0.066) | 0.45 |
| Sex | | |
|   Female | 0.74 (0.065) | 0.47 |
|   Male | 0.74 (0.068) | 0.41 |
| Race | | |
|   Black | 0.71 (0.076) | 0.31 |
|   White | 0.74 (0.064) | 0.60 |
|   Other | 0.74 (0.071) | 0.57 |
| Age | | |
|   0-17 | 0.76 (0.108) | 0.14 |
|   18-39 | 0.59 (0.095) | 0.30 |
|   40-64 | 0.75 (0.068) | 0.54 |
|   65-75 | 0.85 (0.052) | 0.90 |
|   75+ | 0.88 (0.048) | 0.81 |

Table 2. *Comparison between Immunoglobulin G (IgG) to either the nucleocapsid or spike protein (IgG N/S) positivity rates (adjusted by the multilevel regression and poststratification) and administered vaccination rates in the community (Lake and Porter Counties, Indiana) during the week of July 6-12, 2021.*

The MRP IgG prevalence data serve to validate those cohort study findings of prolonged positivity at gradually lower titers, but using a population sample method instead. The coherence of the current study's findings with those cohort studies is a significant reassurance of its validity. Accordingly, our MRP-modified IgG N prevalence curve approximates a plateau from mid-February through June, as can be seen from Figures 1C and 3. As with summer 2020, it was likely that slowly fading IgG N seropositivity during spring 2021 was matched by low level accretion of seropositivity from new infection, hence a roughly flat IgG N curve in both intervals (June-October 2020 in Figure 1C and February-May 2021 in Figure 3).

After we consider both IgG S and IgG N test results, as shown in Figure 2, the MRP modified IgG curve increases rapidly and remains relatively high through time. Our estimates are consistent with the findings in (15) based on 1,443,519 blood donation specimens from a catchment area representing 74% of the U.S. population. The ongoing vaccination process certainly partially explains such an increase and we should be aware that it also suggests the spread of the virus is still not slowing down much, which is also reflected in Table 2.

Table 2 compares the estimated immunity levels, either IgG S or N positive (74%), with the administered vaccination rates in the two counties (45%). Levels of immunity are up in the 59-88% range across all demographic groups, whereas the vaccination rates are highly variable. Black and younger individuals have low vaccination rates, especially in the lowest age category. This observation indicates that Black and youth have acquired substantial immunity through natural exposure to offset their lower vaccination rates. The comparison assumes that the officially released records approximate the vaccination rates of the demographically adjusted groups. Currently we are in the process of attempting to cross reference with respect to the identified vaccination status. Our analysis of the IgG N test results before February 16, 2021 showed that younger age groups had higher tendency to acquire positivity than the elder, but the tendency became lower to have either IgG S or N positive after February 16, 2021.

Throughout the study, we have used IgG N seroprevalence as an estimate for naturally acquired immunity. Once we incorporated IgG S alongside IgG N testing, evidence began to mount that IgG N seropositivity was likely an underestimate of true naturally acquired immunity. A comparison of Table 2 total seropositivity (IgG N or IgG S) with vaccination rates suggests a true natural immunity rate of around 30% in early July (total IgG N/S prevalence minus vaccinated prevalence=29% overall) whereas IgG N positive patients for that interval hover around 5%. Figure 3 shows the estimated IgG N seropositive and IgG S seronegative phenotype prevalence in the hospital system and the community based on asymptomatic patients after February 15, 2021. This finding suggests that around 25% of the population is IgG N negative, but IgG S positive from natural infection. Such patients may have IgG N antibodies below the detection rate of the assay. In any event, for the purposes of this analysis, IgG N positivity should be regarded as merely the lower bound of true naturally-acquired immunity, though the true threshold values for functional immunity are not yet defined.

DISCUSSION

The innovation of our approach lies in the analysis of a demographically stable sample, the development of a pragmatic proxy for random studies measuring population immunity, and the MRP adjustment of demographic discrepancies. In our study, we set up a synthetic random proxy for estimating IgG prevalence in the sampled community. Our model is based on a demographically stable population sample. By applying MRP to our results, we used an accepted statistical modification to correct the otherwise stable demographic skew. We have validated the model using reasonable clinical metrics of viral and symptomatic disease incidence to show the expected biological correlation of these entities with the timing, rate, and magnitude of seroprevalence. In short, the model is able to empower routine hospital testing to measure IgG seroprevalence in the community. We acknowledge that our sampling method will skew slightly in the direction of less healthy individuals than the population as a whole in that it selects for those undergoing outpatient procedures and qualifying for preoperative blood testing.

Of course, IgG positivity is only a single element of overall immunity. Cellular immunity is also clearly crucial, and there is indeed some evidence that humoral and cellular immunity may not fully correlate (16). Nonetheless, while strong humoral immunity may not be required for an effective immune response, it is likely that measurable IgG levels are reasonable evidence of some functional immunity; indeed, high circulating levels of IgG may be the most potent suppressor of disease during a surge in viral transmission. In the event, it is clear that reinfection and vaccine breakthrough do occur, but that preexisting naturally or vaccine-acquired immunity substantially decrease incidence and severity of the disease (17). Thus far, this observation remains valid for current variants of concern as well. In the absence of an easily acquired global view of the immune response within the population, monitoring IgG positivity seems a good way to achieve a sense of that underlying immunity in a policy-actionable sense. As new variants arise and we acquire ongoing experience with their ability to evade naturally and vaccine-acquired humoral immunity, these data will nonetheless continue to be of crucial importance.

Understandably, media and popular scientific conversation on immunity acquisition has been dominated by commentary on the uneven implementation of the vaccine rollout, but this emphasis on vaccine mediated immunity ignores the important humoral immunity demonstrable from the previous infection, as has followed the spring and fall surges in our community and elsewhere. In our data, it is probable that, as of mid-February 2021, 30% or so of our community at large has retained naturally-acquired antibodies to the SARS-CoV-2 nucleocapsid and spike proteins above detection threshold levels, and while the mere presence of such antibodies offered no guarantee of true immunity to second infection within this population, it seems likely that this group largely avoided severe COVID-19 during the delta outbreak, as long as they remained IgG positive. Superimposed vaccination-induced positivity from February onward complicates our ability to discern what percentage of the population may be relying solely on this naturally-conferred immunity persisting. As of the close of the study in July, the presented comparison of publicly available vaccination rates with our normalized measures of IgG seropositivity show that approximately 30% of the population still has assay detectable antibodies obtained by natural infection. Moreover, multiple cohort studies imply that naturally acquired humoral immunity persists for six months or more, so the contribution of this nonvaccinated seropositive group to the total immunity of the community is both significant in magnitude and prolonged in duration. Indeed, delta variant outbreaks in the southern states probably reflected not merely low vaccination rates, but also relatively low prior experience of SARS-CoV-2 infection, in Missouri and Arkansas most notably.

The more recent omicron outbreak presents a distinct challenge to the model. In January 2022, it is becoming increasingly clear that naturally acquired immunity, as well as limited vaccinated immunity, provides less of a barrier to infection and even serious illness than in prior outbreaks. We, therefore, should have less confidence that the immunity identified in our study is reliable as an indicator of community resistance to infection. Indeed, our metric will encounter challenges imposed by the limitations of current testing.

As we develop increasing knowledge of the relative durability and effectiveness of naturally-acquired, vaccine-induced, and combined exposure seropositive status, the power of tracking antibodies arising from natural infection becomes even more fundamental to understanding the total immunity of an identified population to serious infection or death. Combining our metric with state-based data on vaccine administration will allow us a true picture of that overall immunity, informing decisions about opening or closing economic and social commerce within a jurisdiction with more confidence and durability. As we continue to incorporate IgG testing for the spike protein receptor binding domain into our ongoing protocol, the model will be able to follow vaccine acquired immunity as well, avoiding cumbersome additional analysis of state vaccine data, itself often not easily available for analysis due to confidentiality concerns.

Our protocol for serial proxy random sampling of presurgical patients is inexpensive and can be seamlessly integrated into routine preoperative testing in any U.S. hospital. It yields demonstrably valid measures of humoral immunity. Making use of robust demographic data within both hospital EHRs and publicly available census data, this method of sampling allows for a deeper investigation of demographic distributions of immunity and can easily aid in identifying vulnerable populations for targeted vaccine campaigns. Hospital networks and governmental entities can empower the model further by linking up multiple participants in any given jurisdiction via simple universally available database and server constructs. Such a modest effort would further enable confident economic and social policy decisions. Our group is currently developing a Health Insurance Portability and Accountability Act-compliant server-based service to facilitate this broader vision for data access at a statewide or national scale and to ease the data management and interpretation for any given hospital system and its community.

REFERENCES

1. Regenstrief Institute COVID Dashboard. https://www.regenstrief.org/data-downloads/, Accessed October 26, 2021.

2. Gelman A, Little T. Poststratification into many categories using hierarchical logistic regression. Survey Methodology. 1997;23:127–135.

3. Becker M, Strengert M, Junker D, et al. Exploring beyond clinical routine SARS-CoV-2 serology using MultiCoV-Ab to evaluate endemic coronavirus cross-reactivity. Nature Communications. 2021;12(1152).

4. Covello L, Gelman A, Si Y, Wang S. Routine hospital-based SARS-CoV-2 testing outperforms state-based data in predicting clinical burden. Epidemiology. 2021;32(6):792–799.

5. Si Y, Trangucci R, Gabry J, Gelman A. Bayesian hierarchical weighting adjustment and survey inference. Survey Methodology. 2020;46 (2):181–214.

6. Downes M, Gurrin L, English D, Pirkis J, Currier D, Spittal M, Carlin J. Multilevel regression and poststratification: A modeling approach to estimating population quantities from highly selected survey samples. Am J Epidemiol. 2018;187(8):1780–1790.

7. Zhang X, Holt J, Lu H, Wheaton A, Ford E, Greenlund K, Croft J. Multilevel regression and poststratification for small-area estimation of population health outcomes: A case study of chronic obstructive pulmonary disease prevalence using the behavioral risk factor surveillance system. Am J Epidemiol. 2014;179(8):1025-33.

8. Lauer S, Grantz K, Bi Q, et al. The incubation period of Coronavirus Disease 2019 (COVID-19) from publicly reported confirmed cases: Estimation and application. Ann. Int Med. 2020;172(9):577–582.

9. Kissane E, Malaty-Rivera J. Test positivity in the US is a mess. The COVID Tracking Project at The Atlantic. Available at https://covidtracking.com/analysis-updates/test-positivity-in-the-us-is-a-mess. Accessed October 13, 2020.

10. Indiana State Department of Health. COVID-19 region-wide test, case, and death trends. https://hub.mph.in.gov/dataset/covid-19-region-wide-test-case-and-death-trends. Accessed December 21, 2020.

11. Gelman A, Carpenter B. Bayesian analysis of tests with unknown specificity and sensitivity. Journal of the Royal Statistical Society Series C (Applied Statistics). 2020;69(5):1269-1283.

12. Iyer A, et al. Persistence and decay of human antibody responses to the receptor binding domain of SARS-CoV-2 spike protein in COVID-19 patients. Science Immunology. 2020;5(52): eabe0367.

13. Duysburgh E, Mortgat L, Barbezange C, Dierick K, Fischer N, Heyndrickx L, Hutse V, Thomas I, Van Gucht S, Vuylsteke B, Ariën K, Desombere I. Persistence of IgG response to SARS-CoV-2. The Lancet. Infectious diseases. 2021;21(2):163–164.

14. Glück V, Grobecker S, Tydykov L. et al. SARS-CoV-2-directed antibodies persist for more than six months in a cohort with mild to moderate COVID-19. Infection. 2021;49:739–746.

2315. Jones J, Stone M, Sulaeman H, et al. Estimated US infection- and vaccine-induced SARS-CoV-2 seroprevalence based on blood donations, July 2020-May 2021. JAMA. 2021;326(14):1400–1409.

16. Amanna I, Slifka, M. Contributions of humoral and cellular immunity to vaccine-induced protection in humans. Virology. 2011;411(2):206–215.

17. Fontanet A, Cauchemez, S. COVID-19 herd immunity: where are we? Nat Rev Immunol. 2020;20:583–584.

APPENDIX

We poststratified to two different populations: patients in the hospital database (those who have historically and currently obtained care in our regional hospital system) and residents of Lake/Porter County, Indiana. For the hospital, we used the electronic health record (EHR) database in 2019 to represent the population of patients from three hospitals in the Community Health System. For the community, we used the American Community Survey 2014-2018 data from the two counties.

We fit the following logistic regression to allow the PCR prevalence $\pi_i^{PCR}$ for individual $i$ changing over time in the multilevel model parameters.

$$logit(\pi_i^{PCR}) = \beta_1 + \beta_2 male_i + \beta_{age[i]}^{age} + \beta_{race[i]}^{race} + \beta_{county[i]}^{county} + \beta_{time[i]}^{time} + \beta_{age*male[i]}^{age*male}, \quad (1)$$

where $male_i$ is an indicator taking on the value 0.5 for males and -0.5 for females; $age[i], race[i]$, and $county[i]$ represent age, race, and county categories; $age * male[i]$ is the two-way interaction between age and sex; $time[i]$ indices the time in weeks when the test result is observed for individual $i$; and the vectors of varying intercepts are assigned with hierarchical priors:

$$\beta^{name} \sim normal(0, \sigma^{name}), \quad \sigma^{name} \sim normal_+(0, 2.5),$$

for $name \in \{age, race, county, age * male\}$. To allow for the possibility of large variations across time, we set the time-varying effect: $\beta^{time} \sim normal(0, \sigma^{time})$, $\sigma^{time} \sim normal_+(0, 5)$, where $normal_+(a, b)$ represents a half-normal distribution restricted to positive values with the mean value of $a$ and the standard deviation of $b$. The larger the estimated variation, the larger effects of the predictors.

For the IgG model, considering the differences between the two antibody tests, we separated the data before and after February 16, 2021 and fitted the following model of the immunity prevalence $\pi_i^{IgG}$ to the two datasets, respectively:

$$logit(\pi_i^{IgG}) = \alpha_1 + \alpha_2 male_i + \alpha_{age[i]}^{age} + \alpha_{race[i]}^{race} + \alpha_{county[i]}^{county} + \alpha_{time[i]}^{time} + \alpha_{age*time[i]}^{age*time}, \quad (2)$$

where we included an interaction term between age and week indicators to account for the potential vaccination scheduling effect for different age groups. Also, we introduced an informative prior on the interaction term to shrink small interaction effects toward zero:

$$\alpha^{age*time} \sim normal(0, \sigma^{age*time}), \quad \sigma^{age*time} \sim normal_+(0,1).$$

For both the PCR and IgG models, we assume the prior information for the unknown sensitivity $\delta$ and specificity $\gamma$ includes: $y_\gamma$ negative results in $n_\gamma$ tests of known negative subjects and $y_\alpha$ positive results from $n_\alpha$ tests of known positive subjects. The model for the number of positive results $y$ out of $n$ tests is specified as

$$y_\gamma \sim Binomial(n_\gamma, \gamma), \quad y_\delta \sim Binomial(n_\delta, \delta).$$

According to the PCR test protocol, the sensitivity is around 70%, and the specificity is around 100%. We solicit prior information from previous testing results (1). For the sensitivity, the prior data $y_\delta/n_\delta$ are: 70/100, 78/85, 27/37, and 25/35; and the prior data for the specificity $y_\gamma/n_\gamma$ are: 0/0, 368/371, 30/30, 70/70, 1102/1102, 300/300, 311/311, 500/500, 198/200, 99/99, 29/31, 146/150, 105/108, and 50/52. For the IgG tests, considering the estimates of the lower bound, we assume that both the sensitivity and specificity are around 100%.

After obtaining the predicted prevalence from the Bayesian model, we adjusted for selection bias by applying the demographic distributions in the hospital system and the community to generate the population level prevalence estimates, as the poststratification step in MRP. For each of the $2*5*3*2$ cells in the cross-tabulation table of sex (2 levels), age (5 levels), race (3 levels) and county (2 levels), we have the cell-wise incidence estimate $\hat{\pi}_j$, and population count $N_j$, where $j$ is the cell index, and calculate the weekly PCR/IgG prevalence estimate in the population, $\pi_{avg} = \sum_j N_j \hat{\pi}_j / \sum_j N_j$.

REFERENCES

1. Bendavid, E, Mulaney, B, Sood, N, and et al. COVID- 19 antibody seroprevalence in Santa Clara County, California, Version 1. https://www.medrxiv.org/content/10.1101/2020.04.14.20062463v2.full.pdf. Accessed November 21, 2020.